\documentclass[twocolumn,showpacs,showkeys]{revtex4}
\usepackage{graphicx}

\newcommand{\bastar}{\begin{eqnarray*}}
\newcommand{\eastar}{\end{eqnarray*}}
\newskip\humongous \humongous=0pt plus 1000pt minus 1000pt

\newif\ifdtup

\relax
\newcommand{\be}{\begin{equation}}
\newcommand{\ee}{\end{equation}}
\newcommand{\bea}{\begin{eqnarray}}
\newcommand{\eea}{\end{eqnarray}}
\newcommand{\X}{{\vec X}}
\newcommand{\pro}{\partial}
\newcommand{\n}{\hat n}
\newcommand{\oneg}{\displaystyle\frac{1}{g}}

\newcommand{\D}{{\hat D}}

\newcommand{\vX}{{\vec X}}

\newcommand{\vA}{{\vec A}}
\newcommand{\A}{{\vec A}}
\newcommand{\valpha}{{\vec \alpha}}

\newcommand{\hD}{{\hat D}}
\newcommand{\hn}{\hat n}

\newcommand{\tC}{{\tilde C}}
\newcommand{\tD}{\tilde D}
\newcommand{\dfrac}{\displaystyle\frac}
\newcommand{\ba}{\begin{array}}
\newcommand{\ea}{\end{array}}

\newcommand{\nn}{\nonumber}

\begin{document}
\title{Monopole Condensation in $SU(2)$ QCD}
\bigskip
\author{Y.M. Cho}
\email{ymcho@yongmin.snu.ac.kr}
\affiliation{School of Physics, College of Natural Sciences, 
Seoul National University,
Seoul 151-742, Korea\\}
\author{D.G. Pak}  
\email{dmipak@mail.apctp.org}
\affiliation{Asia Pacific Center for Theoretical Physics, 
Seoul 130-012, Korea\\ }
\affiliation{Department of Theoretical Physics, 
Tashkent State University, Tashkent
700-095, Uzbekistan}

\begin{abstract}
Based on the gauge independent decomposition of the non-Abelian
gauge field into the dual potential and the valence potential,
we calculate the one loop effective action of $SU(2)$ QCD
in an arbitrary constant monopole background, using the background
field method. Our result provides a  strong evidence for a dynamical
symmetry breaking through the monopole condensation,
which can induce the dual
Meissner effect and establish the confinement of color, in the non-Abelian
gauge theory.
The result is obtained by separating the topological degrees
which describe the non-Abelian monopoles from the dynamical degrees of
the gauge potential, and integrating out all the dynamical degrees of QCD.
\end{abstract}

\maketitle

\section{Introduction}

One of the most outstanding problems
in theoretical physics is the confinement problem in QCD. It has
long been argued that the monopole condensation could explain the
confinement of color through the dual Meissner effect \cite{nambu,cho1}.
Indeed, if one assumes the monopole condensation, one could easily argue that
the ensuing dual Meissner effect guarantees the confinement \cite{cho2,ezawa}.
In this direction
there has been a remarkable progress in the lattice
simulation during the last decade.
In fact the recent numerical simulations have provided an unmistakable
evidence which supports the idea of the magnetic
confinement through the monopole
condensation \cite{kronfeld,stack}. Unfortunately so far there has been few
satisfactory field theoretic proofs of the monopole condensation in QCD.

The purpose of this paper is to re-examine the non-Abelian dynamics
and establish the monopole condensation in QCD from the first principles.
{\it Utilizing a gauge independent parameterization
of the non-Abelian gauge potential
which emphasizes its topological character, we construct the one loop
effective action of QCD in the presence of monopole background
by integrating out all the dynamical degrees of the non-Abelian
potential except the monople background, using the background
field method. Remarkably the effective action
generates a dynamical symmetry breaking made of the monopole condensation,
which stronly indicates that the physical confinement mechanism
in QCD is indeed the magnetic confinement
through the dual Meissner effect}.
Our analysis makes it clear that it is precisely the magnetic moment
interaction of the gluons which was responsible for the asymptotic
freedom that generates the monopole condensation in QCD.
We demonstrate our result
with $SU(2)$ for simplicity, although the result should be applicable to any
non-Abelian gauge theory.

To prove the magnetic confinement it is instructive for us to
remember how the magnetic flux is confined in the superconductor
through the Meissner effect.  In the macroscopic Ginzburg-Landau description
of superconductivity the Meissner effect is triggered by the
effective mass of the electromagnetic potential, which determines
the penetration (confinement) scale of the magnetic flux.
In the microscopic BCS description, this effective mass
is generated by the electron-pair (the Cooper pair) condensation.
This suggests that, for the confinement of the color electric flux,
one needs the condensation of the monopoles.
Equivalently, in the dual Ginzburg-Landau description, one needs the
dynamical generation of the effective mass for the
monopole potential. To demonstrate this one must first identify the
monopole potential, and separate it from the generic
QCD connection, in a gauge independent manner. This can be done
with the ``Abelian'' projection \cite{cho1,cho2},
which provides us a natural reparameterization
of the non-Abelian connection in terms of
the restricted connection (i.e., the dual potential) of the maximal Abelian
subgroup $H$ of the gauge group $G$ and the gauge covariant vector field
(i.e., the valence potential) of the remaining $G/H$ degrees.
With this separation one can show that
the monopole condensation takes place in one loop correction, after one
integrates out all the dynamical degrees of the non-Abelian gauge
potential.

There have been many attempts to prove the monopole condensation
in QCD \cite{savv,ditt}. Unfortunately the effective action of QCD
obtained from these earlier attempts has failed to
establish the desired magnetic condensation,
because the magnetic condensation was unstable.
This instability of the magnetic condensation
has been widely accepted and never been convincingly revoked.
In retrospect there are many reasons why the earlier attempts
have not been so successful. First the attempts to
calculate the effective action of QCD were gauge dependent. In fact
the separation of the magnetic background from the quantum fields
were not gauge independent. So there is no way of knowing
whether the desired magnetic condensation is indeed 
a gauge independent phenomenon. Moreover the origin of the magnetic
background in the earlier attempts was completely obscure,
and could not be associated to the non-Abelian monopoles.
Consequently the magnetic condensation could not be interpreted as
the monopole condensation. But the most serious defect was
the appearence of an imaginary part in the effective action,
which was due to the improper infra-red regularization.
This improper infra-red regularization was the critical
defect which really destroyed the magnetic
condensation in the earlier attempts \cite{savv,ditt}.
In this paper we start from the gauge independent
separation of the monopole background from the quantum fields
in our calculation of the effective action. More importantly
we make a proper infra-red regularization which respects
the causality, and show that the causality makes
our monopole condensation stable.

Recently Faddeev and Niemi have discovered 
the knot-like topological solitons
in the Skyrme-type non-linear sigma model,
and made an interesting conjecture
that the Skyrme-Faddeev action could be interpreted as an effective
action for QCD in the low energy limit~\cite{faddeev1,faddeev2}.
With the effective action at hand we discuss the possible
connection between Skyrme-Faddeev theory and QCD. We show
that indeed the two theories are closely related,
and demonstrate that we can derive
a generalized Skyrme-Faddeev action from the effective action of QCD.

The paper is organized as follows. In Section II we review
the Abelian projection and the gauge independent decomposition
of the non-Abelian potential into the restricted potential
and the valence potential. In Section III we derive
the integral expression of the one loop effective action
of $SU(2)$ QCD in the presence of pure monopole background,
using the background
field method. In Section IV we derive the integral expression
of the effective action for an arbitrary constant
(color) electromagnetic background, which we need to establish
the stability of the monopole condensation.
In Section V we obtain the effective action for the pure
monopole background, and demonstrate
the existence of the monopole condensation which
generates a dynamical symmetry breaking in QCD.
In Section VI we obtain the effective action for pure electric
background, and show that the electric background generates
the pair annihilation of the valence gluons. In Section VII
we demonstrate the stability of the monopole condensation.
We provide three independent arguments (the causality,
the duality, and the perturbative expansion) which support the stability
of the vacuum condensation. In Section VIII we establish a deep connection
between the Skyrme-Faddeev theory and QCD, and derive a generalized
Skyrme-Faddeev action from our effective action as an
effective action of QCD in the infra-red limit. Finally in
Section IX we discuss the physical implications of our results.

\section{Abelian Projection and Valence Gluon: A Review}

Consider $SU(2)$ QCD for simplicity.  A natural way to identify the
monopole potential is to introduce an isotriplet unit vector field
$\n$ which selects the ``Abelian'' direction (i.e., the color charge direction)
at each space-time point, and to
decompose the connection into the restricted
potential (called the Abelian projection) $\hat A_\mu$ which leaves $\n$
invariant and the valence potential $\vec X_\mu$
which forms a covariant vector field \cite{cho1,cho2},
\bea
 & \vec{A}_\mu =A_\mu \n - \oneg \n\times\pro_\mu\n+\X_\mu\nonumber
         = \hat A_\mu + \X_\mu, \nn\\
 &  (\n^2 =1,~~~ \hat{n}\cdot\vec{X}_\mu=0),
\eea
where $
A_\mu = \n\cdot \vec A_\mu$
is the ``electric'' potential.
Notice that the restricted potential is precisely the connection which
leaves $\n$ invariant under the parallel transport,
\bea
\D_\mu \n = \pro_\mu \n + g {\hat A}_\mu \times \n = 0.
\eea
Under the infinitesimal gauge transformation
\bea
\delta \n = - \vec \alpha \times \n  \,,\,\,\,\,
\delta \A_\mu = \oneg  D_\mu \vec \alpha,
\eea
one has
\bea
&&\delta A_\mu = \oneg \n \cdot \pro_\mu \valpha,\,\,\,\
\delta \hat A_\mu = \oneg \D_\mu \valpha  ,  \nn \\
&&\hspace{1.2cm}\delta \X_\mu = - \valpha \times \X_\mu  .
\eea
{\it This shows that $\hat A_\mu$ by itself describes
an $SU(2)$ connection which
enjoys the full $SU(2)$ gauge degrees of freedom. Furthermore
$\vec X_\mu$ transforms covariantly under the gauge transformation.
Most importantly, the decomposition is gauge-independent. Once
the color direction $\hn$ is selected, the
decomposition follows independent of the choice of a gauge}.
Our decomposition, which has recently
become known as the Cho decomposition \cite{faddeev2} or Cho-Faddeev-Niemi
decomposition \cite{lang}, was first introduced
long time ago in an attempt to demonstrate
the monopole condensation in QCD \cite{cho1,cho2}.
But only recently the importance of the decomposition
in clarifying the non-Abelian dynamics
has become appreciated by many authors \cite{faddeev2,lang}.
Indeed it is this decomposition which has played a crucial role to
establish the Abelian dominance
in Wilson loops in QCD \cite{cho00}, and
the possible connection between the Skyrme-Faddeev action and the
effective action of QCD \cite{cho3,cho4}.

To understand the physical meaning of our decomposition notice that
the restricted potential $\hat{A}_\mu$ actually has a dual structure.
Indeed the field strength made of the restricted potential is decomposed as
\begin{eqnarray}
&\hat{F}_{\mu\nu}=(F_{\mu\nu}+ H_{\mu\nu})\hat{n}\mbox{,}\nonumber\\
&F_{\mu\nu}=\partial_\mu A_{\nu}-\partial_{\nu}A_\mu, \nn\\
&H_{\mu\nu}=-\dfrac{1}{g} \hat{n}\cdot(\partial_\mu\hat{n}\times\partial_\nu\hat{n})
=\partial_\mu \tilde C_\nu-\partial_\nu \tilde C_\mu,
\end{eqnarray}
where $\tilde C_\mu$ is the ``magnetic'' potential
\cite{cho1,cho2}. Notice that we can always introduce the magnetic
potential (at least locally section-wise), because $H_{\mu\nu}$ is closed
\bea
\partial_\mu {\tilde H}_{\mu\nu} = 0 ~~~~~~~ ( {\tilde H}_{\mu\nu} =
\dfrac{1}{2} \epsilon_{\mu\nu\rho\sigma} H_{\rho\sigma} ).
\eea
This allows us to  identify the non-Abelian magnetic potential by
\bea
\vec C_\mu= -\frac{1}{g}\hat n \times \partial_\mu\hat n ,
\eea
in terms of which the magnetic field is expressed as
\bea
&\vec H_{\mu\nu}=\partial_\mu \vec C_\nu-\partial_\nu \vec C_\mu+ g
\vec C_\mu \times \vec C_\nu  \nn\\
&=-g \vec C_\mu \times \vec C_\nu = -\dfrac{1}{g}
\partial_\mu\hat{n}\times\partial_\nu\hat{n} \nn\\
&=H_{\mu\nu}\hat n.
\eea

Another important feature of $\hat{A}_\mu$ is that,
as an $SU(2)$ potential, it retains all the essential
topological characteristics of the original non-Abelian potential.
This is because the topological field $\hat{n}$ can naturally
describe the non-Abelian topology $\pi_2(S^2)$
and $\pi_3(S^2)\simeq\pi_3(S^3)$.
Clearly the isolated singularities of $\hat{n}$ defines $\pi_2(S^2)$
which describes the non-Abelian monopoles.  Indeed $\hat A_\mu$
with $A_\mu =0$ and $\hat n= \hat r$ (or equivallently, $\vec C_\mu$
with $\hat n= \hat r$) describes precisely
the Wu-Yang monopole \cite{wu,cho80}.  Besides, with the $S^3$
compactification of $R^3$, $\hat{n}$ characterizes the
Hopf invariant $\pi_3(S^2)\simeq\pi_3(S^3)$ which describes
the topologically distinct vacua \cite{bpst,cho79}.
This tells that the restricted gauge theory made of $\hat A_\mu$
could describe the dual dynamics which should play an essential
role in $SU(2)$ QCD \cite{cho1,cho00,cho5}.

With (1) we have
\bea
\vec{F}_{\mu\nu}=\hat F_{\mu \nu} + \D _\mu \X_\nu -
\D_\nu \X_\mu + g\X_\mu \times \X_\mu,
\eea
so that the Yang-Mills Lagrangian is expressed as
\bea
&{\cal L}=-\dfrac{1}{4} \vec F^2_{\mu \nu } \nn\\
&=-\dfrac{1}{4}
{\hat F}_{\mu\nu}^2 -\dfrac{1}{4} ( \D_\mu \X_\nu -
\D_\nu \X_\mu)^2 \nn\\ 
&-\dfrac{g}{2} {\hat F}_{\mu\nu}
\cdot (\X_\mu \times \X_\nu) 
- \dfrac{g^2}{4} (\X_\mu \times \X_\nu)^2 \nn\\
&+ \lambda(\hat n^2 -1)
+ \lambda_\mu \hat n \cdot \X_mu,
\eea
where $\lambda$ and $\lambda_\mu$ are the Lagrangian multipliers.
From the Lagrangian we have
\begin{widetext}
\bea
&\partial_\mu ( F_{\mu\nu} + H_{\mu\nu} + X_{\mu\nu} ) = - g
\hn \cdot \Big[ \vX_\mu \times (\hD_\mu \vX_\nu - \hD_\nu \vX_\mu) \Big],
\nn
\\
&\hD_\mu ( \hD_\mu \vX_\nu - \hD_\nu \vX_\mu ) =
g ( F_{\mu\nu} + H_{\mu\nu} + X_{\mu\nu} ) \hn \times \vX_\mu.
\eea
\end{widetext}
where
\bea X_{\mu\nu} = g \hn \cdot ( \vX_\mu \times \vX_\nu ).
\eea
Notice that here $\hat n$ has no equation of motion even though
the Lagrangian contains it explicitly. This
implies that it is not a local degrees of freedom, but a
topological degrees of freedom \cite{cho5}. From this
we conclude that the non-Abelian gauge theory can be viewed
as a restricted gauge theory made of the restricted potential,
which has an additional colored source made of the valence gluon.

Obviously the Lagrangian (10) is invariant under the active
gauge transformation (3). But notice that the decomposition
introduces a new gauge symmetry that we call the passive
gauge transformation \cite{cho3,cho5},
\bea
\delta \hn = 0, ~~~~~~~\delta \vec A_\mu = \dfrac{1}{g} D_\mu \vec \alpha,
\eea
under which we have
\bea
&\delta A_\mu = \dfrac{1}{g} \hn \cdot D_\mu \vec \alpha,
~~~~~~~\delta \hat A_\mu = \dfrac{1}{g} (\hn \cdot D_\mu \vec \alpha) \hn, \nn\\
&\delta \vec X_\mu = \dfrac{1}{g} \Big[D_\mu \vec \alpha
-(\hn \cdot D_\mu \vec \alpha) \hn\Big].
\eea
This is because, for a given $\vA_\mu$,
one can have infinitely many different decomposition of (1),
with different $\hat A_\mu$ and $\vec X_\mu$ by choosing
different $\hn$. Equivalently, for a fixed $\hn$, one can have
infinitely many different $\vec A_\mu$ which are gauge-equivalent
to each other. So it must be clear that with our decomposition we
automatically have another type of gauge invariance which
comes from different choices of decomposition. This extra gauge
invariance plays the crucial role in quantizing the theory \cite{cho5}.

Another advantage of the decomposition (1) is that it can actually
``Abelianize'' (or more precisely ``dualize'') the non-Abelian
dynamics \cite{cho1,cho00,cho5}. To see this let 
$(\hat n_1,~\hat n_2,~\hat n)$
be a right-handed orthonormal basis and let
\begin{eqnarray}
&\vec{X}_\mu =X^1_\mu ~\hat{n}_1 + X^2_\mu ~\hat{n}_2\mbox{,} \nn\\
&(X^1_\mu = \hat {n}_1 \cdot \vec X_\mu,~~~X^2_\mu =
\hat {n}_2 \cdot \vec X_\mu)            \nonumber
\end{eqnarray}
and find
\begin{eqnarray}
&\hat{D}_\mu \vec{X}_\nu =\Big[\partial_\mu X^1_\nu-g
(A_\mu+ \tilde C_\mu)X^2_\nu \Big]\hat n_1 \nn\\
&+ \Big[\partial_\mu X^2_\nu+ g (A_\mu+ \tilde C_\mu)X^1_\nu \Big]\hat{n}_2.
\end{eqnarray}
So with
\bea
& B_\mu = A_\mu + \tC_\mu , \nn\\
&X_\mu = \dfrac{1}{\sqrt{2}} ( X^1_\mu + i X^2_\mu ),
\eea
one could express the Lagrangian explicitly in terms of the dual
potential $B_\mu$ and the complex vector field $X_\mu$,
\begin{eqnarray}
&{\cal L}=-\dfrac{1}{4}(F_{\mu\nu}+ H_{\mu\nu})^2
-\dfrac{1}{2}|\hat{D}_\mu{X}_\nu-\hat{D}_\nu{X}_\mu|^2 \nn\\
&+ ig (F_{\mu\nu} + H_{\mu\nu}) X_\mu^* X_\nu \nn\\
&-\dfrac{1}{2} g^2 \Big[(X_\mu^*X_\mu)^2-(X_\mu^*)^2 (X_\nu)^2 \Big],
\end{eqnarray}
where now
\bea
\hat{D}_\mu{X}_\nu = (\partial_\mu + ig B_\mu) X_\nu.  \nonumber
\eea
Clearly this describes an Abelian gauge theory coupled to
the charged vector field $X_\mu$.
But the important point here is that the Abelian potential
$B_\mu$ is given by the sum of the electric and magnetic potentials $A_\mu+
\tilde C_\mu$.
In this form the equations of motion (11) is re-expressed as
\begin{widetext}
\begin{eqnarray}
&\partial_\mu(F_{\mu\nu}+ H_{\mu\nu}+X_{\mu\nu}) = i g X^*_\mu
( D_\mu X_\nu - D_\nu X_\mu ) - i g X_\mu ( D_\mu X_\nu - D_\nu
X_\mu )^* , \nonumber
\\
&\hat{D}_\mu(\hat{D}_\mu X_\nu- \hat{D}_\nu X_\mu)=ig X_\mu
(F_{\mu\nu}+H_{\mu\nu} +X_{\mu\nu}).
\end{eqnarray}
\end{widetext}
where now
\bea X_{\mu\nu} = - i g ( X_\mu^* X_\nu - X_\nu^* X_\mu ) .  \nonumber
\eea
This shows that one can indeed Abelianize the non-Abelian theory
with our decomposition. The remarkable change in this ``Abelian''
formulation is that here the topological field $\hn$ is
replaced by the magnetic potential $\tilde C_\mu$.

But notice that here we have never fixed
the gauge to obtain this Abelian formalism, and one might
ask how the non-Abelian gauge symmetry is realized in this ``Abelian''
theory. To discuss this let
\bea
&\vec \alpha = \alpha_1~\hn_1 + \alpha_2~\hn_2 + \theta~\hn, \nn\\
&\alpha = \dfrac{1}{\sqrt 2} (\alpha_1 + i ~\alpha_2), \nn\\
&\vec C_\mu = - \dfrac {1}{g} \hn \times \partial_\mu \hn
= - C^1_\mu \hn_1 - C^2_\mu \hn_2, \nn\\
&C_\mu = \dfrac{1}{\sqrt 2} (C^1_\mu + i ~ C^2_\mu).
\eea
Then the Lagrangian (17) is invariant not only under
the active gauge transformation (3) described by
\bea
&\delta A_\mu = \dfrac{1}{g} \partial_\mu \theta -
i (C_\mu^* \alpha - C_\mu \alpha^*),
~~~&\delta \tilde C_\mu = - \delta A_\mu, \nn\\
&\delta X_\mu = 0,
\eea
but also under the passive gauge transformation (13) described by
\bea
&\delta A_\mu = \dfrac{1}{g} \partial_\mu \theta -
i (X_\mu^* \alpha - X_\mu \alpha^*), ~~~&\delta \tilde C_\mu = 0, \nn\\
&\delta X_\mu = \oneg \hD_\mu \alpha - i \theta X_\mu.
\eea
This tells that the ``Abelian'' theory not only retains
the original gauge symmetry, but actually has an enlarged (both the active
and passive) gauge symmetries.
{\it But we emphasize that this is not the ``naive'' Abelianization
of the $SU(2)$ gauge theory which one obtains by fixing the gauge.
Our Abelianization is a gauge independent Abelianization. 
Besides, here the Abelian gauge
group here is actually made of $U(1)_e \otimes U(1)_m$, so that
the theory becomes a dual gauge theory} \cite{cho1,cho00,cho5}. This is
evident from (20) and (21).

\section{Monopole Background}

With this preparation
we will now derive the integral expression of the one-loop
effective action in the presence of the pure monopole background
$\vec C_\mu$. To do this we resort to the background field method
\cite{dewitt,pesk}. So we first divide
the gauge potential $\vec A_\mu$ into two parts, the slow-varying
classical part $\vec A^{(c)}_\mu$ and the fluctuating quantum
part $\vec A^{(q)}_\mu$, and identify the the magnetic potential $\vec C_\mu$
as the classical background \cite{cho3,cho5},
\bea
&\vec A_\mu = \vec A^{(c)}_\mu + \vec A^{(q)}_\mu, \nn\\
&\vec A^{(c)}_\mu = \vec C_\mu,~~~~~\vec A^{(q)}_\mu = A_\mu \hat n + \X_\mu.
\eea
With this we introduce two types of gauge transformations, the background
gauge transformation and the physical gauge transformation.
Naturally we identify the background gauge transformation as
\bea
&\delta \vec C_\mu = \dfrac{1}{g} \bar D_\mu \vec \alpha,\nn\\
&\delta (A_\mu \hat n + \X_\mu) = - \vec \alpha \times
(A_\mu \hat n + \X_\mu),
\eea
where now $\bar D_\mu$ is defined with only
the background potential $\vec C_\mu$
\bea
\bar D_\mu = (\partial_\mu + g \vec C_\mu \times).
\eea
As for the physical gauge transformation which leaves the
background potential invariant, we must have
\bea
\delta \vec C_\mu = 0,~~~~~\delta (A_\mu \hat n + \X_\mu) =
\dfrac{1}{g} D_\mu \vec \alpha.
\eea
Notice that both (23) and (25) respect the original
gauge transformation,
\bea
\delta \A_\mu =
\dfrac{1}{g} D_\mu \vec \alpha.
\eea
Now, we fix the gauge by
imposing the following gauge condition to the quantum fields,
\bea
&\vec F = \bar D_\mu (A_\mu \hat n + \vec X_\mu) =0 , \nn\\
&{\cal L}_{gf} =- \dfrac{1}{2\xi}
\left[(\partial_\mu A_\mu)^2 + ({\bar D}_\mu \X_\mu)^2\right].
\eea
The corresponding Faddeev-Popov determinant is given by
\bea
M^{ab}_{FP} = \dfrac {\delta F^a}{\delta \alpha^b} = (\bar D_\mu D_\mu)^{ab}.
\eea
With this gauge fixing
the effective action takes the following form,
\begin{widetext}
\bea
&\exp~\Big[iS_{eff} (\vec C_\mu) \Big]
= \dfrac{}{} \int {\cal D} A_\mu {\cal D}
\X_\mu {\cal D} \vec{c} ~{\cal D}\vec{c}^{~*}
\exp \Big \{{i \dfrac{}{} \int \Big[-\dfrac {1}{4}{\hat F}_{\mu \nu}^2} 
-\dfrac{1}{4} ( \D_\mu \X_\nu -\D_\nu \X_\mu)^2 \nn\\
&-\dfrac{g}{2} {\hat F}_{\mu\nu} \cdot (\X_\mu \times \X_\nu)
-\dfrac{g^2}{4}(\X_\mu \times \X_\nu)^2
+\vec{c}^{~*}\bar {D}_\mu D_\mu\vec{c} 
-\dfrac{1}{2\xi}(\partial_\mu A_\mu)^2-\dfrac{1}{2\xi}
(\bar {D}_\mu\vec{X}_\mu)^2 \Big]d^4x \Big\},
\eea
\end{widetext}
where $\vec c$ and ${\vec c}^{~*}$ are the ghost fields.
Notice that the effectice action (29) is explicitly invariant
under the background gauge transformation (3),
if we add the following
gauge transformation of the ghost fields to (3),
\bea
\delta \vec c = - \alpha \times \vec c,
~~~~~\delta \vec c^{~*} = - \alpha \times \vec c^{~*}.
\eea
This guarantees that the resulting effective action we obtain
after the functional integral should be invariant under
the remaining background gauge trnsformation which involves
only $\vec C_\mu$. This, of course, is the advantage of the
background field method which greatly simplifies the calculation
of the effective action \cite{dewitt,pesk}.

Now, we can perform the functional integral in (29). Remember that in one loop
approximation only the terms quadratic in quantum fields
become relevant in the functional integral.
So the $A_\mu$ integration becomes trivial,
and the $\X_\mu$ and ghost integrations result in the
following functional determinants (with $\xi=1$),
\bea
&{\rm Det}^{-\frac{1}{2}} K_{\mu \nu}^{ab}\simeq
{\rm Det}^{-\frac{1}{2}} \Big[-g_{\mu \nu}
 (\bar D \bar D)^{ab}
- 2gH_{\mu \nu}\epsilon^{abc} n^c \Big], \nn\\
&{\rm Det} M^{ab}_{FP} \simeq {\rm Det} (-\bar{D} \bar{D})^{ab}.
\eea
One can simplify the determinant $K$ \cite{cho3,cho6}
\bea
&\ln {\rm Det}^{-\frac{1}{2}} K 
=-\dfrac12\ln {\rm Det} \Big[(-\bar{D}\bar{D})^{ab}
+i\sqrt{2}gH\epsilon^{abc}n^c \Big]\nn\\
&-\dfrac12\ln {\rm Det} \Big[(-\bar{D}\bar{D})^{ab}
-i\sqrt{2}gH\epsilon^{abc}n^c \Big] \nn\\
&-\ln {\rm Det}(-\bar{D}\bar{D})^{ab},
\eea
where
\bea
H=\sqrt{\vec{H}_{\mu\nu}^2}. \nn
\eea
With this the one loop contribution of the functional
determinants to the effective action can be written as
\bea
&\Delta S = i\ln {\rm Det}(-\bar {D}^{2} +\sqrt{2}gH) \nn\\
&         + i\ln {\rm Det} (-\bar {D}^{2} -\sqrt{2}gH),
\eea
where now $\bar {D}_\mu$ acquires the following Abelian form
\bea
\bar {D}_\mu =\partial_\mu + ig\tilde{C}_\mu .
\eea
Remarkably the functional determinants (33) acquires the Abelian form.
This, of course, is precisely due to the fact that our decomposition
(1) Abelianizes QCD. But we emphasize again that this Abelianization
is gauge independent.

One can evaluate the functional determinants in (33) with the
Fock-Schwinger proper time method, and for a constant background
$H$ we find
\bea
&\Delta{\cal L} = \dfrac{1}{16 \pi^2}\int_{0}^{\infty}
\dfrac{dt}{t^2}
 \dfrac{g H/\sqrt{2} \mu^2}{\sinh (g H t/\sqrt{2}\mu^2)}\nn \\
&\Big[ \exp (-\sqrt{2}g H t/\mu^2 )+ \exp (\sqrt{2}g H t/\mu^2) \Big],
\eea
where $\mu$ is a dimensional parameter.

\section{Arbitrary Background}

Before we evaluate the above integral and establish the monopole condensation
we now derive the integral expression of the one-loop
effective action in the presence of arbitrary background
$\hat A_\mu$, which we need to establish the stability of
the monopole condensation. So we repeat the above procedure,
but now replacing the monopole background $\vec C_\mu$ by
the restricted potential $\hat A_\mu$. So we first divide
the gauge potential $\vec A_\mu$ into two parts,
and now identify the restricted potential $\hat A_\mu$
as the classical background,
\bea
&\vec A_\mu = \vec A^{(c)}_\mu + \vec A^{(q)}_\mu, \nn\\
&\vec A^{(c)}_\mu = \hat A_\mu,~~~~~\vec A^{(q)}_\mu = \X_\mu.
\eea
With this we introduce two types of gauge transformations, the background
gauge transformation and the physical gauge transformation.
Naturally we identify the gauge transformation (3) as
the background gauge transformation.
As for the physical gauge transformation which leaves the
background potential invariant, we must have
\bea
\delta \hat A_\mu = 0,~~~~~\delta \X_\mu =
\dfrac{1}{g} D_\mu \vec \alpha.
\eea
Again notice that both (3) and (37) respect the original
gauge transformation (26).
Now, we fix the gauge by
imposing the following gauge condition to 
the quantum field \cite{cho5,cho6},
\bea
&\vec F = \hat D_\mu \vec X_\mu =0 , \nn\\
&{\cal L}_{gf}=- \dfrac{1}{2\xi}
(\hat D_\mu \vec X_\mu)^2.
\eea
The corresponding Faddeev-Popov determinant is given by
\bea
M^{ab}_{FP} = \dfrac {\delta F^a}{\delta \alpha^b} = (\hat D_\mu D_\mu)^{ab}.
\eea
With this gauge fixing
the effective action takes the following form,
\begin{widetext}
\bea
&\exp ~\Big[iS_{eff}(\hat A_\mu) \Big] = \dfrac{}{} \int {\cal D}
\X_\mu {\cal D} \vec{c} ~{\cal D}\vec{c}^{~*}
\exp \Big\{{~i \dfrac{}{} \int \Big[-\dfrac {1}{4}{\hat F}_{\mu \nu}^2}
-\dfrac{1}{4} ( \D_\mu \X_\nu -\D_\nu \X_\mu)^2 \nn\\
&-\dfrac{g}{2} {\hat F}_{\mu\nu} \cdot (\X_\mu \times \X_\nu)
-\dfrac{g^2}{4}(\X_\mu \times \X_\nu)^2
+\vec{c}^{~*}\hat {D}_\mu D_\mu\vec{c}
-\frac{1}{2\xi}
(\hat {D}_\mu\vec{X}_\mu)^2 \Big] d^4x \Big\}.
\eea
\end{widetext}
Notice again that, with (30), the effectice action (40) is explicitly invariant
under the background gauge transformation (3) which involves
only $\hat A_\mu$.

Now, we can perform the functional integral. To do this let
the background field $\hat F_{\mu\nu}$ be
\bea
&\hat F_{\mu\nu}= G_{\mu\nu}\hat n, \nn\\
&G_{\mu\nu} = F_{\mu\nu} +H_{\mu\nu}.\nn
\eea
Since, in one loop approximation only the terms which are quadratic in
the quantum fields are relevant to the functional integral,
we find that the $\X_\mu$ and ghost integrations result in the
following functional determinants (with $\xi=1$),
\bea
&{\rm Det}^{-\frac{1}{2}} K_{\mu \nu}^{ab}\simeq 
{\rm Det}^{-\frac{1}{2}} \Big[-g_{\mu \nu}
 (\hD \hD)^{ab}- 2g G_{\mu \nu}\epsilon^{abc} n^c \Big],\nn \\
& {\rm Det} M_{FP} = {\rm Det} \Big[-(\hat D \hat D)^{ab} \Big] ,
\eea
where now ${\hat D}_\mu$ is defined with an arbitrary
background field ${\hat A}_\mu$.
Using the relation
\bea
&G_{\mu \alpha} G_{\nu\beta} G_{\alpha \beta} = \dfrac{1}{2} G^2 G_{\mu \nu}
+\dfrac{1}{2}(G \tilde G) {\tilde G}_{\mu \nu} \nn\\
&({\tilde G}_{\mu \nu}=\dfrac{1}{2}{\epsilon}_{\mu\nu\rho\sigma}
G_{\rho\sigma}),
\eea
one can simplify the functional determinants of the valence
gluon and the ghost loops to the following Abelian form,
\bea
& \ln {\rm Det}^{-\frac 1 2} \Big[(- g_{\mu \nu} (\hat D \hat D)^{ab}
-2g G_{\mu\nu} \epsilon ^{abc} n^c \Big]  \nn\\
& = \ln {\rm Det} \Big[(-\tD^2+2a)(-\tD^2-2a) \nn\\
&(-\tD^2-2ib)(-\tD^2+2ib) \Big],\nn\\
& \ln {\rm Det}M_{FP} = 2\ln {\rm Det}(-\tD^2),
\eea
where  now $\tilde D_\mu$ is defined with an arbitrary
background $A_\mu + {\tilde C}_\mu$,
\bea
\tD_\mu = \pro +ig(A_\mu + {\tilde C}_\mu),
\eea
and
\bea
&a = \dfrac{g}{2} \sqrt {\sqrt {G^4 + (G \tilde G)^2} + G^2}, \nn\\
&b = \dfrac{g}{2} \sqrt {\sqrt {G^4 + (G \tilde G)^2} - G^2}.
\eea
Notice that in the Lorentz frame where the
electric field becomes parallel to the magnetic field, $a$ becomes
purely magnetic and $b$ becomes purely electric.

From this we have
\bea
&\Delta S = i\ln {\rm Det} \Big[(-\tD^2+2a)(-\tD^2-2a) \Big] \nn\\
&+ i\ln {\rm Det} \Big[(-\tD^2-2ib)(-\tD^2+2ib) \Big] \nn\\
& - 2i\ln {\rm Det}(-\tD^2).
\eea
We can evaluate the functional determinants, and for a general
background with arbitrary $a$ and $b$, the contribution
of the gluon and ghost loops is given by \cite{cho6}
\bea
&\Delta {\cal L} =  \dfrac{1}{16 \pi^2}
\int_{0}^{\infty} \dfrac{ d t}{t^3}
\dfrac{ a b t^2/\mu^2}{\sinh  (a t/\mu^2) \sin (b t/\mu^2) }\nn \\
& \Big[\exp(-2at/\mu^2)+\exp(2at/\mu^2) \nn\\
&+\exp(2ibt/\mu^2)+\exp(-2ibt/\mu^2)-2 \Big].
\eea
The integral expression
(47) of the effective action has been
known for some time \cite{ditt}, but the actual integration
of it is not easy to perform. Indeed, as far as we understand,
the integration has never been evaluated correctly.
This is because the integral contains (not only the usual
ultra-violet divergence around $t \simeq 0$) a severe infra-red divergence
around $t \simeq \infty$, which has to be regularized correctly.
In the following we will perform the integral for pure magnetic
and pure electric backgrounds separately.

\section{Monopole Condensation}

For the pure monopole background
the integral (35) reduces to
\begin{widetext}
\bea
&\Delta {\cal L}=  \Delta{\cal L_+} + \Delta{\cal L_-}, \nn\\
& \Delta{\cal L_+} =  \dfrac{1}{16 \pi^2}\int_{0}^{\infty}
\dfrac{dt}{t^2}\dfrac{a/\mu^2}{\sinh(at/\mu^2)}
\exp(-2at/\mu^2) \nn \\
&\Delta{\cal L_-} =   \dfrac{1}{16 \pi^2}\int_{0}^{\infty}
\dfrac{dt}{t^2} \dfrac{a/\mu^2}{\sinh(at/\mu^2)}
\exp(2at/\mu^2),
\eea
\end{widetext}
where
\bea
a=\dfrac{gH}{\sqrt{2}}. \nn
\eea
Notice that this is precisely the same integral that we obtain from
(47) for the pure magnetic background (i.e., for $b=0$).
This tells that the evaluation of
the effective action for an arbitrary magnetic background becomes
mathematically identical to that for the pure monopole background.

As we have remarked both integrals
have the usual ultra-violet divergence at the origin, but the second
integral has a severe infra-red divergence at the infinity.
To find the correct infra-red regularization, one must understand
the origin of the divergence. The infra-red divergence
can be traced back to the
magnetic moment interaction of the gluons that we have in (10), which
is also well-known to be responsible for the asymptotic
freedom \cite{gross}. This magnetic interaction
generates negative eigenvalues in Det K in the long distance
region, which cause the infra-red divergence.
More precisely when the momentum $k$ of the gluon
parallel to the background magnetic field becomes smaller than
the background field strength (i.e., when $k^2 < a$), the lowest
Landau level gluon eigenfunction whose spin is parallel to
the magnetic field acquires an imaginary energy and thus becomes tachyonic.
It is these unphysical tachyonic states which cause the infra-red
divergence. So one must exclude these tachyonic modes in the calculation
of the effective action, when one makes a proper infra-red regularization.
Including the tachyons in the physical spectrum will surely destablize
QCD and make it ill-defined.

The correct infra-red regularization is dictated by the causality.
To implement the causality in (48) we first go to the Minkowski time
with the Wick rotation, and find
\begin{widetext}
\bea
& \Delta{\cal L_+} =  - \dfrac{1}{16 \pi^2}\int_{0}^{\infty}
\dfrac{dt}{t^2} \dfrac{a /\mu^2}{\sin (at/\mu^2)} 
\exp (-2i a t/\mu^2 ), \nn\\
& \Delta{\cal L_-} =  - \dfrac{1}{16 \pi^2}\int_{0}^{\infty}
\dfrac{dt}{t^2} \dfrac{a/\mu^2}{\sin (at/\mu^2)} 
\exp (+2i a t/\mu^2 ).
\eea
\end{widetext}
In this form the infra-red divergence has disappeared,
but now we face the ambiguity in choosing the correct contours
of the integrals in (49). Fortunately this ambiguity can
be resolved by the causality. To see this notice that the two integrals
$\Delta{\cal L_+}$ and $\Delta{\cal L_-}$ originate from the
two determinants in (33), and the standard causality argument requires us to
identify $2 a$ in the first determinant as
$2 a -i\epsilon$ but in the second determinant as
$2 a +i\epsilon$. {\it This tells that
the poles in the first integral in (49) should lie above the real
axis, but the poles in the second integral should lie
below the real axis. From this we conclude
that the contour in $\Delta{\cal L_+}$ should pass below the
real axis, but the contour in $\Delta{\cal L_-}$ should pass above the
real axis. With this causality requirement the two integrals
become complex conjugate to each other, which guarantees that
$\Delta{\cal L}$ is explicitly real, without any imaginary part}.
This removes the infra-red divergence. We emphasize that this
causality for the infra-red regularization is precisely the same causality
that determines the Feynman propagators in field theory.
With this observation we finally have
\bea
&\Delta{\cal L} = \dfrac{1}{16 \pi^2}\int_{0}^{\infty}
\dfrac{ d t}{t^{2-\epsilon}}
\dfrac{a/\mu^2}{\sinh  (a t/\mu^2 ) } \nn\\
& \Big[ \exp (-2 a t/\mu^2 )+  \exp (-2 a t/\mu^2 ) \Big],
\eea
where now $\epsilon$ is the ultra-violet cutoff which we have introduced
to regularize the ultra-violet divergence.

Now we can perform the integral, and obtain
\bea
&\Delta{\cal L} = \dfrac{11a^2}{48\pi^2}(\dfrac{1}{\epsilon}-\gamma)
-\dfrac{11a^2}{48\pi^2}(\ln\dfrac{a}{\mu^2}-c), \nn\\
&c=1-\ln 2 -\dfrac {24}{11} \zeta'(-1, \frac{3}{2})=0.94556... ,
\eea
where $\zeta(x,y)$ is the generalized Hurwitz zeta function.
So with the ultra-violet regularization by modified minimal subtraction
we finally obtain the following effective Lagrangian \cite{cho3,cho6}
\bea
{\cal L}_{eff}=-\dfrac{1}{2g^2}a^2 -\dfrac{11}{48\pi^2}a^2(\ln
\dfrac{a}{\mu^2}-c ).
\eea
This completes our derivation of the one-loop effective Lagrangian of $SU(2)$
QCD in the presence of the monopole background. Notice that,
as expected, the effective Lagrangian is explicitly invariant
under the background gauge transformation (23) which involves only
$\vec C_\mu$.

\begin{figure*}
\includegraphics{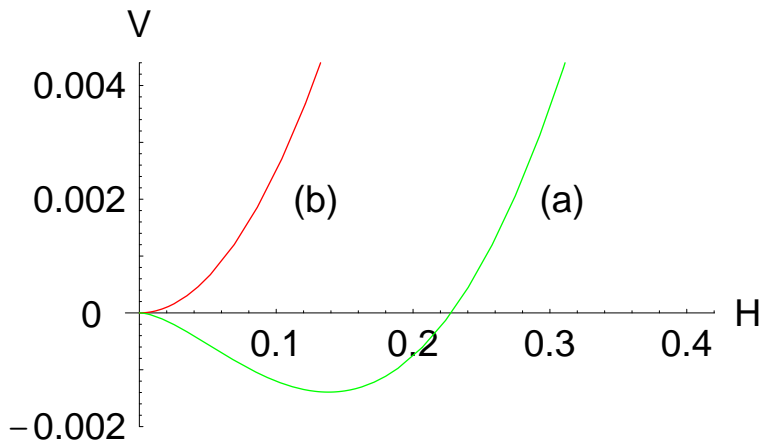}
\caption{\label{Fig. 1} The effective potential of SU(2)
QCD in the pure magnetic background.
Here (a) is the effective potential and (b) is the classical potential.}
\end{figure*}

As we have indicated, there is another way to obtain
the effective action which is more physical.
Remember that the two integrals in (48) come from the two determinants
in (33), and the infra-red divergence in the second integral
comes from the tachyonic modes contained in the
second determinant in (33). So one can calculate the effective action
by calculating the determinant correctly. Now, to evaluate the determinant
one is supposed to use a complete set of
eigenfunctions which is made of the physical states.
But obviously the tachyonic modes can not be regarded as
physical, because
they violate the causality. {\it A remarkable point is that
by calculating the determinants with the physical states
one can show that the second determinant become identical to
the first one. This means that, by calculating
the functional determinants correctly, one can
obtain exactly the same effective action that we have obtained
with the infra-red regularization by causality}. This provides
another justification of
our effective action (52).

Now we are ready to establish the monopole
condensation. To do this we renormalize the effective
action first. For this notice that
the effective action provides the following non-trivial
effective potential
\bea
V=\frac14 H^2\Big[1+\frac{11 g^2}{24 \pi^2}(\ln\frac{gH}{\mu^2}-c_1)\Big],
\eea
where
\bea
c_1 = 1 - \dfrac{1}{2} \ln 2 - \dfrac {24}{11} \zeta'(-1, \frac{3}{2})=
1.29214...\, . \nn
\eea
So we can define the running coupling $\bar g$ by \cite{savv}
\bea
\frac{\partial^2V}{\partial H^2}\Big|_{H=\bar \mu^2}
=\frac{1}{2}\frac{g^2}{ \bar g^2}.
\eea

With the definition we obtain
\bea
\frac{1}{\bar g^2} =
\frac{1}{g^2}+\frac{11}{24 \pi^2}( \ln\frac{g {\bar\mu}^2}{\mu^2}
- c_1 + \dfrac{3}{2}),
\eea
from which we obtain the following $\beta$-function,
\bea
\beta(\bar\mu)=-\frac{11}{24\pi^2} \bar g^3~.
\eea
{\it This is exactly the same $\beta$-function that one obtained
from the perturbative QCD to prove the asymptotic freedom}
\cite{gross}. This confirms that our effective action is
consistent with the asymptotic freedom.

{\it The fact that the $\beta$-function obtained from
the effective action becomes identical to the one obtained by
the perturbative calculation is really remarkable, because
this is not always the case. In fact in QED it has been
demonstrated that the running coupling and the $\beta$-function
obtained from the effective action is different from those
obtained from the perturbative method} \cite{cho01,cho7}.

In terms of the running coupling the renormalized potential is given by
\bea
V_{\rm ren}=\frac14 H^2\Big[1+\dfrac{11}{24 \pi^2 } \bar g^2
(\ln\dfrac{H}{\bar\mu^2}-\dfrac{3}{2})\Big],
\eea
which generates a non-trivial local minimum at
\bea
<H>=\frac{\bar \mu^2}{\bar g} \exp\Big(-\frac{24\pi^2}{11\bar g^2}+ 1\Big).
\eea
Notice that with ${\bar \alpha}_s = 1$ we have
\bea
    \dfrac{<H>}{{\bar \mu}^2} = 0.13819... .
\eea
{\it This is nothing but the desired magnetic condensation.
This proves that the one
loop effective action of QCD in the presence of the constant magnetic
background does generate a dynamical symmetry breaking thorugh the
monopole condensation} \cite{cho3,cho6}.

The corresponding effective potential is plotted in Fig.1,
where we have assumed $\bar \alpha_s = 1, ~\bar \mu =1$.
The effective potential clearly shows that there is indeed a dynamical
symmetry breaking in QCD.

The renormalization group invariance of the effective action
is guaranteed by the Callan-Symanzik equation
\bea
\Big(\bar\mu\frac{\partial}{\partial \bar\mu}+\beta\frac{\partial}{\partial \bar g}
-\gamma( \vec C_\mu)\vec{C}_\mu\frac{\partial}{\partial\vec{C}_\mu} \Big)V_{\rm ren}=0,
\eea
where $\gamma (\vec C_\mu)$ is the anomalous dimension for $\vec C_\mu$,
\bea
\gamma( \vec C_\mu)=-\frac{11}{48\pi^2}\bar g^2+O(\bar g^4).
\eea
This should be compared with that of the gluon field in perturbative QCD,
$\gamma(\vec{A}_\mu)=-5\bar g^2/24\pi^2$ for $SU(2)$, in the absence
of the quarks.

\section {Electric Background}

To make sure that our infra-red regularization is indeed the correct one
it is necessary to have an independent confirmation of the above result.
To do this it is instructive to calculate the effective action
with a pure electric background first.

From (46) and (47) we have for a pure electric background (i.e., for $a=0$)
\bea
&\Delta S = i\ln {\rm Det}(-\tD^2-2ib) \nn\\
&+ i\ln {\rm Det}(-\tD^2+2ib),
\eea
and
\bea
&\Delta {\cal L}  =  \dfrac{1}{16 \pi^2}  \int_{0}^{\infty}
\dfrac{ d t}{t^2}
\dfrac{b/\mu^2}{\sin (bt/\mu^2) } \nn\\
& \Big[\exp(2ibt/\mu^2)+\exp(-2ibt/\mu^2) \Big].
\eea
There are different ways to evaluate the integral, but a simple
and nice way of doing this follows from the observation that in
the imaginary time (i.e., in the Minkowski time) the role of
the electric and magnetic fields are reversed. So with the Wick rotation
the above integral acquires
the same form as (48). Indeed with the Wick rotation (63) becomes
\bea
&\Delta {\cal L} = -\dfrac{1}{16 \pi^2} i^{\epsilon}
\int_{0}^{\infty} \dfrac{ d t}{t^{2-\epsilon}}
\dfrac{b}{\sinh(bt) } \nn\\
& \Big[\exp(-2bt)+\exp(2bt) \Big].
\eea
Now, adopting the same infra-red regularization as in
the pure magnetic background, we obtain
\bea
&\Delta{{\cal L}}=
-\dfrac{11b^2}{48\pi^2}(\dfrac{1}{\epsilon}-\gamma)
+\dfrac{11b^2}{48\pi^2}(\ln\dfrac{b}{\mu^2}-c) \nn\\
&-i\dfrac{11b^2}{96\pi}.
\eea
So with the modified minimal subtraction we have (with the pure
electric background)
\bea
{\cal L}_{eff} = \dfrac{b^2}{2g^2} +\dfrac{11b^2}{48\pi^2}
(\ln \dfrac{b}{\mu^2}-c)-i\dfrac{11b^2}{96\pi}.
\eea
We emphasize that in evaluating
the above integral the same infra-red regularization is applied
as in the pure magnetic background.
With the pure electric background
the eigenfunctions of the second determinant in (62)
becomes anti-causal and thus unphysical in the long
distance region (i.e., for $k^2 < b$),
just like the eigenfunctions under the pure magnetic background
become tachyonic and unphysical in the infra-red
region (i.e., for $k^2 < a$).
So we must again exclude these
unphysical modes to evaluate the above integral.

Another way to perform the integral (63) is by choosing the proper
contour. Notice that (unlike the pure magnetic background) the integrand
here has poles on the real axis, so that we must
specify the contour of the integral. To find the proper contour,
first notice that the eigenvalues of the two determinants in (62)
are complex conjugate to each other. This means that
the contour of the two integrals in (63) should also be complex
conjugate to each other. Secondly, one can make the first integral finite
by choosing the contour to pass above the real axis and rotating
it to the positive imaginary axis (i.e., by replacing $t$ with
$it$). This is justifiable, because the first integral is free
of the controversial acausal states. With this the contour of
the second integral is fixed by complex conjugating the first
contour. This means that the second contour must pass below
the real axis, which one can rotate to the
negative imaginary axis (by replacing $t$ with $-it$).
This makes the second integral finite. Finally the causality
requires us to replace $b$ with $b+\epsilon$ in the first
determinant but $b-\epsilon$ in the second determinant in (62).
This means that the first contour should start from $0+\epsilon$,
but the second one from $0-\epsilon$ in (63). From this we conclude that
the half of the residue at the origin should contribute to
the integral. This recipe reproduces (66),
and justifies the result.

Notice that it is the causality
that produces the imaginary part in (65). This is remarkable,
because it was the same causality which has made (51)
explicitly real. So in both pure magnetic and pure electric
backgrounds the causality determines the imaginary part
of the effective action.

The contrast between the effective actions (52) and (66)
is remarkable. First, the effective potential derived from (66)
has no local minimum. This implies that the electric background
does not generate a condensation. Secondly, (66) has an imaginary part
\bea
Im \thinspace {\cal L} =-\dfrac{11b^2}{96\pi}.
\eea
This implies that the electric background is unstable. {\it But
perhaps a more important point here is that the imaginary part is negative.
This means that the electric background generates the pair annihilation,
rather than the pair creation, of the gluons. This is because the negative
imaginary part can be interpreted as the negative probability of
the pair creation. This implies that the gluons in QCD, unlike
the electrons in QED, tend to annihilate among
themselves in the color electric field.}
This might sound strange, but actually is not difficult to
understand.  Indeed this is precisely what the asymptotic freedom
dictates. To understand this remember that the gluon loop
contributes positively, but the quark loop contributes 
negatively, to the asymptotic freedom \cite{gross}. 
Exactly for the same reason
the gluon and quark loops contribute oppositely to the 
imaginary part of the effective action. But  
the quark loop in QCD, just like
the electron loop in QED, generates a positive
imaginary part \cite{cho6}. This tells that the gluon loop
should generate a negative imaginary part in the effective
action. This means that the asymptotic freedom, 
the anti-screening, and the pair 
annihilation all originate from the same physics. This is 
really remarkable.

\section{Stability of Monopole Condensation}

There have been many attempts to construct the effective action of QCD
in the literature, and in the appearance our vacuum (58)
looks very much like the old Savvidy-Nielsen-Olesen
(SNO) vacuum \cite{savv,ditt}.
The major difference is that the effective action in
the earlier approaches contained an imaginary part, which
made the magnetic condensation unstable.  In contrast our
effective action is explicitly real, which guarantees the stability
of our monopole condensation. Indeed
it has been asserted that the SNO
vacuum should be unstable, because the effective action
which defines the vacuum develops an imaginary part
\cite{savv,ditt},
\bea
Im \thinspace {\cal L} \Big |_{SNO} = ~\dfrac {1} {8\pi} a^2.
\eea
This destabilizes the vacuum through the pair creation of gluons.
This assertion of the instability of the
SNO vacuum, which comes from
improper infra-red regularizations, has been widely accepted and
never been convincingly revoked.
As a consequence it has been generally believed that the one-loop
effective action can not establish the monopole condensation in QCD.
Our analysis tells that this misleading belief has no foundation.

But since the absence of the absorptive part in our effective action
is such a crucial point which distinguishes our effective action
from the SNO action, one might
like to have an independent proof that our infra-red regularization
is indeed the correct one.
Fortunately there are various ways to make an independent confirmation
of our effective actions (52) and (66). To see this first notice that
the imaginary part (68) of the SNO action as well as
ours are quadratic in the background fields.
This, with the definition (45), tells that
the imaginary part of the one loop effective action is second
order in the coupling constant $g$. So one can find the
correct imaginary part of the effective action perturbatively,
just by calculating the effective action up to
the second order in the coupling constant in the perturbative
expansion. There are different ways of doing this.
In fact one can just calculate the relevant Feynmann
diagrams of the perturbative expansion \cite{schan},
or can adopt the Schwinger's method used in QED
to obtain the imaginary part \cite{schw}.
Now, a remarkable point is that these perturbative calculations do
reproduce the result which is identical to ours \cite{cho8},
\bea
Im \thinspace \Delta {\cal L}=\left\{{~~~~0~~~~~~~~~~~~~~ b=0~~
\atop -\dfrac{11b^2}{96\pi}~~~~~~~~~~~a=0~.}\right.
\eea
{\it This confirms that our infra-red regularization is indeed correct.
More importantly this confirms that we do have the desired dynamical
symmetry breaking and the magnetic condensation in QCD.}
It must be pointed out that the possibility that
one could calculate the imaginary part of the effective
action by the perturbative method, and that the SNO action
could probably be incorrect, was first raised by
Schanbacher \cite{schan}. Unfortunately this remarkable work has been
completely neglected so far, probably because this work is
also plagued by the defect that it is not gauge independent.

We emphasize that this perturbative calculation of the imaginary part
in QCD is justified precisely because the imaginary part of
the effective action is second order in $g$. This is remarkable,
because in general the one loop effective action does not
allow a perturbative expansion. For example in QED
the perturbative expansion of the imaginary
(as well as the real) part of the effective action is divergent
and does not make sense, because the point $e=0$ is singular
\cite{cho01,cho7}. This means that in QED the perturbative calculation does
not reproduce the result of one loop effective action.

To reinforce our assertion we now provide a third
independent argument which supports our results.
{\it An important point to observe here is that the effective actions (52)
and (66) are actually the mirror image of each other.
To see this notice that we can obtain
(66) from (52) simply by replacing $a$ with $-ib$, and similarly
(52) from (66) by replacing $b$ with $ia$.
This is the first indication
that there exists a fundamental symmetry which we call the duality
in the effective action of QCD. The duality states that
the effective action must be invariant under the replacement}
\bea
a \rightarrow - ib,~~~~~~~b \rightarrow ia. \nn
\eea
This type of duality was first eastablished
in the effective action of QED \cite{cho01,cho7}. But 
we emphasize that exactly the same duality should also
hold in our effective action of QCD, because we have already
Abelianized it.
An important point of the duality is that the duality provides
a very useful tool to check the consistency of the effective action.
In the present case the duality indeed confirm the consistency of
our effective actions (52) and (66). Obviously this endorses
that our calculation of the imaginary parts (69) is probably
correct, or at least consistent with the duality.
This tells that the causality,
the perturbative expansion, and the duality all strongly
endorse the stability of our monopole condensation.

It must be emphasized that there are fundamental differences between the
earlier attempts and the present approach.
The earlier attempts had three problems.
First the separation between
the classical background and the quantum field was not gauge independent,
which made it difficult to establish the gauge invariance of
the one loop effective action.
Secondly the origin of the magnetic background
has never been clarified. As a consequence the magnetic
condensation could not be associated with the monopole
background. These defects were
serious enough, but perhaps the most serious
problem was that the infra-red divergence was not properly regularized
in the earlier attempts. Because of this the SNO effective
action contained an imaginary part.
This destabilizes the vacuum through the pair creation of gluons.

{\it In contrast in our approach the separation of the monopole background
from the quantum fluctuation is clearly gauge independent.
Moreover our infra-red regularization
generates no imaginary part in the effective action. Because of these
we obtain a stable vacuum made of monopole condensation which is
both gauge and Lorentz invariant}.
Notice that the infra-red regularization in (50) is not just to remove the
infra-red divergence (there are infinitely many ways to do this). The infra-red
divergence that we face here in QCD is also different from those one encounters
in the effective action of the massless QED \cite{cho01,cho7}.
The infra-red divergence in the massless QED
comes from the zero modes. But these zero modes are physical modes,
which should not be excluded in the calculation of the effective
action. On the other hand the infra-red divergence
that we have here comes from the unphysical modes,
so that one must exclude these unphysical modes from the physical spectrum with
a proper infra-red regularization. Our analysis has provided ample
reason why this has to be so (Notice that in the earlier attempts
these tachyonic modes are incorrectly identified as
the ``unstable'' modes, but we emphasize
that they are not just unstable but unphysical).
And it is precisely these unphysical modes
that generate the controversial imaginary part in
the SNO action.
So with the exclusion of the unphysical modes the instability
of the vacuum disappears completely.
As importantly in our approach we can really
claim that the magnetic condensation is a gauge independent phenomenon.
Furthermore here we have demonstrated that it is precisely
the Wu-Yang monopole that is responsible for the
condensation.

\section {QCD versus Skyrme-Faddeev Action}

Recently Faddeev and Niemi have discovered the knot-like topological solitons
in the Skyrme-type non-linear sigma model~\cite{faddeev1},
\bea
&{\cal
L}_{SF} = - \dfrac{\mu^2}{2} (\partial_\mu \hat n)^2 -
\dfrac{1}{4} (\partial_\mu \hat n \times \partial_\nu \hat n)^2,
\eea
and made an interesting conjecture
that the Skyrme-Faddeev action could be interpreted as an effective
action for QCD in the low energy limit~\cite{faddeev2}. But we emphasize
that from our decomposition (1) it should have been evident that the above
action is closely related to QCD. Indeed
from the decomposition we have \cite{cho4}
\bea
{\cal L}_{SF} = -
\dfrac{1}{4} \vec H_{\mu\nu}^2 - \dfrac{\mu^2} {2} \vec C_\mu^2.
\eea
This tells that the Skyrme-Faddeev theory can be interpreted
as a massive Yang-Mills theory where the gauge potential has the
special form (7). Furthermore we can claim that it is a theory of
monopoles and at the same time a theory of confinement, where the
monopole-anti-monopole pairs are confined 
to form the knots \cite{cho4,cho5}.
But now with the effective action of QCD at hand
we can discuss
the connection between QCD and Skyrme-Faddeev theory in more detail.

Evidently the effective action (52) is invariant under
both gauge and Lorentz transformations. On the other hand
we can express the effective action explicitly
in terms of the monopole field strength
$\vec H_{\mu\nu}$. This, of course,
is not accidental. The background field method
guarantees that the effective action should be expressed by the
gauge invariant form, invariant under the background gauge transformation (23).
What is remarkable here is that, with (7), the background
magnetic field $\vec H_{\mu\nu}$ can be expressed completely
by the magnetic potential $\vec C_\mu$
\bea
\vec H_{\mu\nu} = -g \vec C_\mu \times \vec C_\nu, \nn
\eea
so that the effective potential (53) can actually
be written completely in terms of $\vec C_\mu$,
\bea
&V=\dfrac{g^2}{4}(\vec{C}_\mu\times\vec{C}_\nu)^2\Big \{ 1 \nn\\
&+\dfrac{11 g^2}{24 \pi^2}
\Big[ \ln \dfrac{g[(\vec{C}_\mu\times\vec{C}_\nu)^2]^{1/2}}
{\mu^2}-c_1\Big]\Big\}.
\eea
Now, just for a heuristic reason, suppose we choose
a particular Lorentz frame and
express the vacuum (58) by the vacuum expectation value of $\vec C_\mu$.
In this case the above effective potential generates the following
mass matrix for $\vec C_\mu$,
\bea
{\displaystyle M}^{ij}_{\mu\nu} = \Big< \dfrac {\delta^2 V}
{\delta {C}^{i}_{\mu} \delta {C}^{j}_{\nu}} \Big>
= m^2~(\delta^{ij} - n^i n^j) g_{\mu\nu},
\eea
where
\bea
m^2=\frac{11 g^4}{96\pi^2}\Big<\frac{(\vec{C}_\mu\times
\vec{H}_{\mu\nu})^2}{H^2}\Big>
\eea
can be interpreted as the ``effective mass'' for $\vec C_\mu$.
This demonstrates that the magnetic condensation indeed generates the
mass gap necessary for the dual Meissner effect and
the confinement.

With the above understanding we can now study the possible connection
between the Skyrme-Faddeev action and the effective actin of QCD.
To do this we first expand the effective potential in terms of
the monopole potential around the vacuum and make the following
Taylor expansion,
\begin{widetext}
\bea
&V = V_0 + \dfrac {1}{2!} {\Big <} \dfrac {\delta^2 V}
{\delta C^i_\mu \delta C^j_\nu}{\Big >}
~{\bar C}^i_\mu ~{\bar C}^j_\nu
+ \dfrac{1}{3!}  {\Big <} \dfrac {\delta^3 V}{\delta C^i_\mu \delta
C^j_\nu \delta C^k_\rho}{\Big >}
~{\bar C}^i_\mu ~{\bar C}^j_\nu ~{\bar
C}^k_\rho \nn \\
&+ \dfrac{1}{4!}  {\Big <} \dfrac {\delta^4 V}{\delta C^i_\mu \delta
C^j_\nu \delta C^k_\rho \delta C^l_\sigma }{\Big >}
~{\bar C}^i_\mu ~{\bar C}^j_\nu ~{\bar C}^k_\rho ~{\bar C}^l_\sigma + .........,
\eea
\end{widetext}
where
\bea
{\bar C}^i_\mu = C^i_\mu - < C^i_\mu>. \nn
\eea
Now, near the vacuum we could neglect the higher order terms and
keep only the quartic polynomial in $\vec C_\mu$
for simplicity. In this approximation the corresponding
effective Lagrangian will acquire the form
\begin{widetext}
\bea
&{\cal L}_{eff} = - \dfrac12m^2(\vec{C}_\mu) ^2
- \dfrac{\alpha}{4} (\vec{C}_\mu\times\vec{C}_\nu)^2
- \dfrac{\beta}{4} (\vec C_\mu \cdot \vec C_\nu)^2 - \dfrac{\gamma}{4}
(\vec C_\mu)^4
+ ..........\nn\\
&=-\dfrac{m^2}{2g^2}(\partial_\mu \hat{n})^2
-\dfrac{\alpha}{4g^2}(\partial_\mu \hat{n}
\times\partial_\nu \hat{n})^2
- \dfrac{\beta}{4g^2} (\partial_\mu \hat n \cdot \partial_\nu \hat n)^2
- \dfrac{\gamma}{4g^2} (\partial_\mu \hat n)^4 + .........,
\eea
\end{widetext}
where $\alpha$, $\beta$, and $\gamma$ are numerical parameters
which can be fixed from (75). This is nothing but
a generalized Skyrme-Faddeev Lagrangian \cite{cho3,cho4}.
This shows that one can indeed derive a generalized Skyrme-Faddeev action
from QCD by expanding the effective potential around the vacuum.
This, together with (71), establishes a firm connection between
the Skyrme-Faddeev theory and QCD. In fact we can go further, and
establish a deep connection between QCD and the Skyrme
theory itself \cite{cho4,cho5}.

An important feature in our analysis is that the Skyrme-Faddeev action
is intimately connected to the monopole condensation in QCD.
In particular our analysis makes it clear that the mass scale
in the Skyrme-Faddeev action is
directly related to the mass of the monopole potential,
which determines the confinement scale in QCD. This is not surprizing.
Indeed any attempt to relate the Skyrme-Faddeev action to QCD
must produce the mass scale that the Skyrme-Faddeev action contains,
and the only way to interpret this mass scale in QCD is through
the confinement.

But it must be emphasized that our approximation (76) is
by no means exact. There are two points that should
be kept in mind here. First, we have
kept only the quadratic part and neglected all the higher order
terms in (76). More seriously, in deriving the effective action we
have neglected the derivatives of $\vec H_{\mu\nu}$ and
thus the derivatives of $\vec C_\mu$, assuming that
$H$ is constant.
Secondly, we had to choose a particular Lorentz frame to justify
the expansion (75) of the effective action around the vacuum.
So our derivation appears to have compromised the Lorentz invariance,
although the generalized Skyrme-Faddeev action is obviously Lorentz invariant.
Consequently our analysis establishes a possible connection
between a ``generalized'' non-linear sigma model of Skyrme-Faddeev type
and QCD only in a limited sense. In particular it does not assert
that the simple-minded Skyrme-Faddeev action
describes QCD in the infra-red limit.
In spite of these drawbacks our analysis strongly
endorses the fact that the Skyrme-Faddeev action
has something in common with QCD, which is really remarkable.

\section{Discussion}

In this paper we have established the monopole condensation,
which describes a stable vacuum of QCD. Furthermore
we have demonstrated the existence of a genuine
dynamical symmetry breaking in QCD triggered by
the monopole condensation. We were
able to do this by calculating the one loop effective action of $SU(2)$
QCD in the presence of pure monopole background. There have
been earlier attempts to calculate the effective
action, but our result differs from the earlier results. The main
difference with the earlier attempts was the controversial imaginary
part in the effective action in the earlier attempts.
This has made the SNO vacuum unstable.
In contrast, with a proper
infra-red regularization, we have shown that the QCD vacuum
made of the monopole condensation is stable.
We have provided three
independent arguments to support our conclusion.

It is truly remarkable that the principles of the quantum
field theory allow us to demonstrate the monopole condensation
within the framework of the conventional quantum field
theory. The assertion of the instability of the SNO vacuum
has created a wrong impression that one can not demonstrate
the monopole condensation with the one-loop effective action.
Our analysis tells that in truth one can demonstrate the
monopole condensation with the effective action.
Notice, however, that this does
not prove that the monopole condensation is the true vacuum
of QCD. To prove this we have to calculate the effective action
in an arbitrary color electromagnetic background, and show that indeed
the monopole condensation is the true minimum of the effective
potential. This is not an easy task. Even for the ``simple'' QED
the calculation of the one loop
effective action in an arbitrary background has been completed
only recently \cite{cho01,cho7}, fifty years after the Schwinger's seminal
work \cite{schw}. In the subsequent paper we obtain
the one loop effective action of QCD for an arbitrary
background, and demonstrate that indeed the monopole
condensation is the true vacuum of QCD, at least at one loop level
\cite{cho6,cho9}.

We conclude with the following remarks: \\
1) It should be emphasized that the
gauge independent decomposition (1) of the non-Abelian gauge potential
plays the crucial role in our analysis.
The decomposition
has been known for more than twenty years \cite {cho1,cho2},
but its physical significance
appears to have been appreciated very little till recently.
Now we emphasize that it is this decomposition
which has made the gauge independent separation of the classical
background from the quantum field, and allows us
to obtain the effective action of QCD without ambiguity.
In particular, it is this decomposition which shows that the vacuum
condensation is indeed made of the monopole condensation.
Many of the earlier approaches had the critical defect that the
decomposition of the non-Abelian gauge potential to the $U(1)$ potential
and the charged vector field
was not gauge independent, which has made these approaches 
controversial. In particular, in these approaches one can not make sure 
that the effective action (and the resulting magnetic condensation) 
obtained with the Abelian
background really has a gauge independent meaning. \\
2) There have been two competing proposals for the correct mechanism
of the confinement in QCD, the one emphasizing the role of the instantons and
the other emphasizing that of the monopoles. Our analysis strongly
favors the monopoles as the physical
source for the confinement. It provides a natural dynamical symmetry
breaking, and generates the mass
gap necessary for the confinement in QCD.
Notice that the multiple  vacua, even though it is an important
characteristics of the non-Abelian gauge theory, 
did not play any crucial role
in our calculation of the effective action. 
Moreover our result shows that it is
the monopole condensate, not the $\theta$-vacuum, which describes
the physical vacuum of QCD. \\
3) We have established a firm connection between the
Skyrme-Faddeev action and QCD. On the other hand the Skyrme-Faddeev
theory (and the Skyrme theory itself) contains the topological knot
states. If so, QCD could also likely to admit such states, which
might naturally be interpreted as the ``glueballs''.
But these knots are not the ordinary glueballs made of the valence
gluons. They are made of the magnetic, not electric, flux.
In this sense they should be called the ``magnetic'' glueballs \cite{cho4}.
The existence of such magnetic glueballs has been
predicted long time ago \cite{cho1,cho2}. Once the monopole condensation
sets in, one should expect
the fluctuation of the condensed vacuum.  But obviously the fluctuation
modes have to be magnetic, which could be identified as
the magnetic glueballs (A new feature here is that they
have a topological stability. But this could be an artifact of
the effective theory, not a genuine feature of QCD). We can even
predict that the mass of these glueballs starts from 
around 1.4 GeV \cite{cho4}.
If so, the remaining task is to look for a convincing experimental evidence
of the magnetic glueball states in hadron spectrum \cite{cho1,cho2}.

Although we have concentrated to $SU(2)$ QCD in this paper, it must be clear
from our analysis that the magnetic condensation is a generic
feature of the non-Abelian gauge theory.
A more detailed discussion which contains the calculation
of the effective action in the presence of an arbitrary
color electromagnetic background
will be presented in an accompanying paper \cite{cho9}.\par

{\bf Acknowledgements}

~~~One of the authors (YMC) thanks S. Adler and F. Dyson
for the fruitful discussions, and Professor C. N. Yang for
the continuous encouragements. The other (DGP)
thanks Professor C. N. Yang for the fellowship at Asia Pacific
Center for Theoretical Physics, and appreciates Haewon Lee
for numerous discussions.
The work is supported in part by Korea Research Foundation (Grant KRF-2001
-015-BP0085) and by the BK21 project of Ministry of Education.


\begin{thebibliography}{99}

\bibitem{nambu}Y. Nambu, Phys. Rev. {\bf D10}, 4262 (1974);
S. Mandelstam, Phys. Rep. {\bf 23C}, 245 (1976);
A. Polyakov, Nucl. Phys. {\bf B120}, 429 (1977);
G. 't Hooft, Nucl. Phys. {\bf B190}, 455 (1981).
\bibitem{cho1}Y. M. Cho, Phys. Rev. {\bf D21}, 1080 (1980);
J. Korean Phys. Soc. {\bf17}, 266 (1984).
\bibitem{cho2}Y. M. Cho, Phys. Rev. Lett. {\bf 46}, 302 (1981);
Phys. Rev. {\bf D23}, 2415 (1981).
\bibitem{ezawa}Z. Ezawa and A. Iwazaki, Phys. Rev. {\bf D25}, 2681 (1982);
T. Suzuki, Prog. Theor. Phys. {\bf 80}, 929 (1988);
H. Suganuma, S. Sasaki, and H. Toki, Nucl. Phys. {\bf B435}, 207 (1995);
K. Kondo, Phys. Rev. {\bf D57}, 7467 (1998); {\bf D58}, 105016 (1998).
\bibitem{kronfeld}A. Kronfeld, G. Schierholz, and U. Wiese, Nucl. Phys.
{\bf B293}, 461 (1987);
T. Suzuki and I. Yotsuyanagi, Phys. Rev. {\bf D42}, 4257 (1990).
\bibitem{stack}J. Stack, S. Neiman, and R. Wensley, Phys. Rev. {\bf D50},
3399 (1994);
H. Shiba and T. Suzuki, Phys. Lett. {\bf B333}, 461 (1994);
G. Bali, V. Bornyakov, M. M\"{u}ller-Preussker, and K. Schilling,
Phys. Rev. {\bf D54}, 2863 (1996).
\bibitem{savv} G. K. Savvidy, Phys. Lett. {\bf B71}, 133 (1977);
N. K. Nielsen and P. Olesen, Nucl. Phys. {\bf B144}, 485 (1978);
C. Rajiadakos, Phys. Lett. {\bf B100}, 471 (1981).
\bibitem{ditt} A. Yildiz and P. Cox, Phys. Rev. {\bf D21}, 1095 (1980);
M. Claudson, A. Yilditz, and P. Cox, Phys. Rev. {\bf D22}, 2022 (1980);
W. Dittrich and M. Reuter, Phys. Lett. {\bf B128}, 321, (1983);
C. Flory, Phys. Rev. {\bf D28}, 1425 (1983);
S. K. Blau, M. Visser, and A. Wipf, Int. J. Mod. Phys.
{\bf A6}, 5409 (1991); M. Reuter, M. G. Schmidt, and C. Schubert,
Ann. Phys. {\bf 259}, 313 (1997).
\bibitem{faddeev1} L. Faddeev and A. Niemi, Nature {\bf 387}, 58 (1997);
R. Battye and P. Sutcliffe, Phys. Rev. Lett. {\bf 81}, 4798 (1998).
\bibitem{faddeev2} L. Faddeev and A. Niemi, Phys. Rev. Lett.
{\bf 82}, 1624 (1999); Phys. Lett. {\bf B449}, 214 (1999).
\bibitem{lang}E. Langman and A. Niemi, Phys. Lett. {\bf B463}, 252 (1999);
S. Shabanov, Phys. Lett. {\bf B458}, 322 (1999); {\bf B463}, 263 (1999);
H. Gies, hep-th/0102026.
\bibitem{cho00} Y. M. Cho, Phys. Rev {\bf D62}, 074009 (2000).
\bibitem{cho3} Y. M. Cho and D. G. Pak, J. Korean Phys.
Soc. {\bf 38}, 151 (2001); Y. M. Cho, H. W. Lee, and D. G. Pak, 
hep-th/9905125, Phys. Lett. {\bf B525}, 347 (2002).
\bibitem{cho4} Y. M. Cho, Phys. Rev. Lett. {\bf 87}, 252001 (2001).
\bibitem{wu}T. T. Wu and C. N. Yang, Phys. Rev. {\bf D12}, 3845 (1975).
\bibitem{cho80} Y. M. Cho, Phys. Rev. Lett. {\bf 44}, 1115 (1980);
Phys. Lett. {\bf B115}, 125 (1982); Y. M. Cho and D. Maison, Phys. Lett.
{\bf B391}, 360 (1997).
\bibitem{bpst}A. Belavin, A. Polyakov, A. Schwartz, and Y. Tyupkin,
Phys. Lett. {\bf B59}, 85 (1975); G. 't Hooft, Phys. Rev. Lett.
{\bf 37}, 8 (1976).
\bibitem {cho79} Y. M. Cho,
Phys. Lett. {\bf B81}, 25 (1979).
\bibitem{cho5} W. S. Bae, Y. M. Cho, and S. W. Kimm, 
Phys. Rev. {\bf D65}, 025005 (2002).
\bibitem{dewitt} B. De Witt, Phys. Rev. {\bf 162}, 1195 (1967);
{\it ibid}, 1239 (1969); J. Honerkamp, Nucl. Phys. {\bf B48}, 269 (1972).
\bibitem{pesk} See for example,
C. Itzikson and J. Zuber, {\it Quantum Field Theory} (McGraw-Hill) 1985;
M. Peskin and D. Schroeder,
{\it An Introduction to Quantum Field Theory} (Addison-Wesley) 1995;
S. Weinberg, {\it Quantum Theory of Fields} (Cambridge Univ. Press) 1996. 
\bibitem{cho6} Y. M. Cho and D. G. Pak,
in {\it Proceedings of TMU-YALE
Symposium on Dynamics of Gauge Fields}, edited by T. Appelquist and H.
Minakata (Universal Academy Press) Tokyo, 1999.
\bibitem{gross} D. Gross and F. Wilczek, Phys. Rev. Lett. {\bf 26}, 1343 (1973);
H. Politzer, Phys. Rev. Lett. {\bf 26}, 1346 (1973).
\bibitem {cho01} Y. M. Cho and D. G. Pak, Phys. Rev. Lett. {\bf 86},
1947 (2001); W. S. Bae, Y. M. Cho, and D. G. Pak, Phys. Rev. {\bf D64},
017303 (2001).
\bibitem{cho7} Y. M. Cho and D. G. Pak, hep-th/0010073, submitted
to Phys. Rev. Lett.
\bibitem{schan} V. Schanbacher, Phys. Rev. {\bf D26}, 489 (1982).
\bibitem{schw} J. Schwinger, Phys. Rev. {\bf 82} , 664 (1951).
\bibitem{cho8} Y. M. Cho and M. Walker, hep-th/000000, submitted
to Phys. Rev. {\bf D}.
\bibitem{cho9} Y. M. Cho and D. G. Pak, hep-th/0006051, submitted
to Phys. Rev. {\bf D}.
\end{thebibliography}
\end{document}